\documentclass[10pt, aps,prc,twocolumn,superscriptaddress,preprintnumbers,
amsmath, 
floatfix,
longbibliography,
nofootinbib
]{revtex4-1}
\usepackage[T1]{fontenc}
\usepackage[utf8x]{inputenc} 
\usepackage{adjustbox}          
\usepackage[caption=false]{subfig}
\usepackage{url}
\usepackage{color}
\usepackage{float}
\usepackage[pdftex,colorlinks=true, linkcolor = blue, citecolor=blue,urlcolor=blue, bookmarksnumbered=true, bookmarksopen=true]{hyperref}
\usepackage{longtable}
\usepackage{amsfonts}
\usepackage{wrapfig,bm,bbm}
\newcommand{\beq}{\begin{equation}}
\newcommand{\eeq}{\end{equation}}
\newcommand{\bea}{\begin{eqnarray}}
\newcommand{\eea}{\end{eqnarray}}

\begin{document}

\title{Fragment Intrinsic Spins and Fragments' Relative Orbital Angular Momentum in Nuclear Fission} 

\author{Aurel Bulgac}
\affiliation{Department of Physics, University of Washington, Seattle, Washington 98195--1560, USA}
\author{Ibrahim Abdurrahman} 
\affiliation{Department of Physics, University of Washington, Seattle, Washington 98195--1560, USA}
\author{Kyle Godbey}
\affiliation{Facility for Rare Isotope Beams, Michigan State University, East Lansing, Michigan 48824, USA}
\author{Ionel Stetcu}
\affiliation{Theoretical Division, Los Alamos National Laboratory, Los Alamos, New Mexico 87545, USA}

\date{\today}

\begin{abstract}

We present the first fully unrestricted microscopic calculations of the primary fission fragment intrinsic spins and of the 
fission fragments' relative orbital angular momentum for $^{236}$U$^*$, 
$^{240}$Pu$^*$, and $^{252}$Cf  using the time-dependent density functional theory framework.  Within this microscopic approach, 
free of restrictions and unchecked assumptions and which incorporates the relevant physical observables for describing fission, 
we evaluate the triple distribution of the fission fragment intrinsic spins  and of 
their fission fragments' relative orbital angular momentum and show that  their dynamics is dominated by their bending collective 
modes, in contradistinction to the predictions of the existing phenomenological models and some interpretations of experimental data. 

\end{abstract}

\preprint{NT@UW-21-08, LA-UR-21-27848}

\maketitle

While nuclear fission has been studied for more than eight decades~\cite{Hahn:1939},
a complete microscopic description based on quantum many-body theory is still lacking.
Typical microscopic approaches rely on unverified assumptions and/or strong restrictions, thus rendering the treatment  incomplete.
Phenomenological models are based on the imagination of their creators, rather than rigorous quantum mechanics 
or direct experimental information.
\textcite{Meitner:1939} correctly identified the main driver of nuclear fission: namely, the competition between the 
Coulomb energy and the surface potential energy. The formation of the compound nucleus 
and its extremely slow shape evolution toward the outer fission barrier is 
correctly encapsulated by Bohr's compound nucleus concept~\cite{Bohr:1936,Bohr:1939}. 
The saturation properties of nuclei along with the symmetry energy constrain the flow of the nuclear fluid
from the moment the compound nucleus is formed until scission, which evolves like an incompressible 
liquid drop of almost constant local proton-neutron mixture. 
The spin-orbit interaction and pairing correlations
control the finer details on how the emerging fission fragments (FFs) are formed,  
favoring asymmetric fission yields at low excitation energies~\cite{Strutinsky:1967,BRACK:1972,Bertsch:1980,Bertsch:1997}. 
The critical theoretical ingredients are thus well-known: the incompressibility of nuclear matter, the symmetry energy strength, 
the surface tension and the proton charge, the spin-orbit and the pairing correlations strengths. 
Only recently, a well-founded formalism free of restrictions that incorporates all of these features has been implemented and
the nonequilibrium character of the nuclear large amplitude collective motion, 
particularly from the outer saddle to the scission configuration and the excitation energy sharing mechanism between FFs 
have been unambiguously proven microscopically~\cite{Bulgac:2016,Bulgac:2019b,Bulgac:2020}.

The FFs' intrinsic spins have been the subject of old and renewed experimental and theoretical investigations
 \cite{Strutinsky:1960,Ericson:1960,Nix:1965,Moretto:1980,Wilson:2021,Bulgac:2021,Marevic:2021,Randrup:2021}. 
In the 1960s, it was conjectured that the emerging FFs acquire intrinsic spins due 
to the existence of several collective FF spin modes: the double-degenerate transversal modes, wriggling and bending, 
and the longitudinal modes, twisting and tilting. 
The origin of the relative orbital angular momentum between fragments has never been elucidated within a fully microscopic
framework.  Consider the clean case of spontaneous fission of $^{252}$Cf 
from its ground state with $S_0^\pi = 0^+$.   
The final three angular momenta satisfy the conservation law
\begin{align} 
{\bm S_0} = {\bm S}^\text{L} + {\bm S}^\text{H} +{\bm \Lambda} ={\bm 0} \quad \text{in case of}\quad  ^{252}\text{Cf} , \label{eq:S0}
\end{align}
where ${\bm S}^\text{L,H}$ are the FF intrinsic spins and ${\Lambda}$ is the FFs' relative orbital angular momentum, 
which is an integer. Classically, these three vectors lie in a plane and ${\bm \Lambda}={\bm R}\times{\bm P}$,  
is perpendicular to the fission direction, where ${\bm R},{\bm P}$ are the FFs' relative separation and momentum.
On its way to scission this nucleus elongates along a spontaneously broken symmetry direction 
and the fledging FFs emerge. The longer the nuclear elongation the larger the moment of inertia of 
the entire nuclear system is and the overall rotational frequency controlled by ${\bm \Lambda}$ is slower.  
As FFs emerge, being by nature nonspherical, 
they rotate with intrinsic spins ${\bm S}^\text{L}$ and  ${\bm S}^\text{H}$, 
while at the same time they also rotate as a dumbbell around their common center of mass with the angular momentum ${\bm \Lambda}$.  
Until scission, these three angular momenta can vary, subject to  restriction Eq.~\eqref{eq:S0}. After scission, when the mass and energy 
exchange between emerging FFs stops, these angular momenta cease to evolve in time (apart from small effects of the Coulomb 
interaction between FFs~\cite{Strutinsky:1960,Bulgac:2020a}). Before scission the FF identities are not well-defined, 
because matter, momentum, and energy are flowing between them. The FF intrinsic spins and ${\bm \Lambda}$ are 
well-defined only at a sufficiently relative large separation.  Even though the initial nuclear system $^{252}$Cf 
has a vanishing initial spin $S^\pi_0=0^+$, the FFs emerge as 
wave packets of deformed nuclei, characterized by rotation and vibrational bands.  Similar
to the well-known bicycle wheel classroom physics demos~\cite{Phys-demo}, the dynamics of a 
spontaneously fissioning $^{252}$Cf resembles the dynamics of an instructor 
on a freely rotating stand (${\bm  \Lambda}$)  holding two bicycle wheels (${\bm S}^\text{L,H}$), and is nothing like a 
``snapping rubber band''~\cite{Wilson:2021}, which does not rotate.   

We use the time-dependent density functional theory (TDDFT) extended to superfluid systems (see recent 
reviews~\cite{Bulgac:2013a,Bulgac:2019} and Refs.~\cite{Bulgac:2016,Bulgac:2019b,Bulgac:2020,Bulgac:2021})
to determine the triple probability distribution $P(S^\text{L},S^\text{H},\Lambda)$, \,
$\sum_{S^\text{L,H},\Lambda} P(\Lambda,S^\text{L},S^\text{H})=1$, 
by performing a triple
angular momenta projection of the overlap~\cite{Bulgac:2021c}  
\begin{align}
\!\!\!\!\!\langle \Phi | \Phi(\beta_0,\beta_\text{L},\beta_\text{H}) \rangle  =
\langle \Phi | e^{i\beta_0(J_x^\text{L}+J_x^\text{H})}  e^{i\beta^\text{L} J_x^\text{L}}e^{i\beta^\text{H}J_x^\text{H}}  |\Phi\rangle . \label{eq:over}
\end{align}
where $Oz$ the fission axis and the magnitudes of the angular momenta satisfy the triangle restriction
\begin{align}
|{S}^\text{L}-{S}^\text{H}| \le {\Lambda} \le { S}^\text{L}+{ S}^\text{H} \label{eq:3Js}
\end{align}
and $|\Phi\rangle$ is the fissioning nucleus intrinsic wave function.
In case of $^{236}$U$^*$ and $^{240}$Pu$^*$  the initial spin $S_0\neq 0$ and then $|{\bm \Lambda}-{\bm S}_0| = |{\bm S}^\text{L}+{\bm S}^\text{H}|$ and
since  $S_0 \ll \langle \Lambda\rangle$ then $\Lambda \approx |{\bm S}^\text{L}+{\bm S}^\text{H}|$ with good accuracy.  
We determined the probability distribution $p(\cos\phi^\text{LH})$, where $\phi^\text{LH}$ is the angle between 
$S^\text{L}$ and $S^\text{H}$ by constructing a histogram  of the expectation of the cosine between 
\begin{align}
\cos \phi^\text{LH} = \frac{ \Lambda(\Lambda+1) - S^\text{L}(S^\text{L}+1) -S^\text{H}(S^\text{H}+1)}{2(S^\text{L}+1/2)(S^\text{H}+1/2)}, \label{eq:cos}
\end{align}
where we used the Langer correction~\cite{Langer:1937} in the denominator. 
Note that the relative angle $\phi^\text{LH}$ does not  depend on a lab or body reference frame. 
Optimally, one should consider also an additional  projection to enforce the value of total angular momentum 
${\bm S}_0$, with the rotation operator $P_0 =  e^{i\gamma(J_x^\text{L}+J_x^\text{H}+ \Lambda_x)}$, 
where $\Lambda_x$ rotates the entire system around its center of mass, a procedure that is expected 
to lead only to minor corrections~\cite{Bulgac:2021}. We replaced this projection with the equivalent triangle restriction 
\begin{align}
\triangle  = \Theta({\Lambda}\ge |{S}^\text{L}-{S}^\text{H}|) \Theta({ \Lambda}\le {S}^\text{L}+{ S}^\text{H}).\label{eq:Delta}
\end{align} 

\begin{figure}
\includegraphics[width=1\columnwidth]{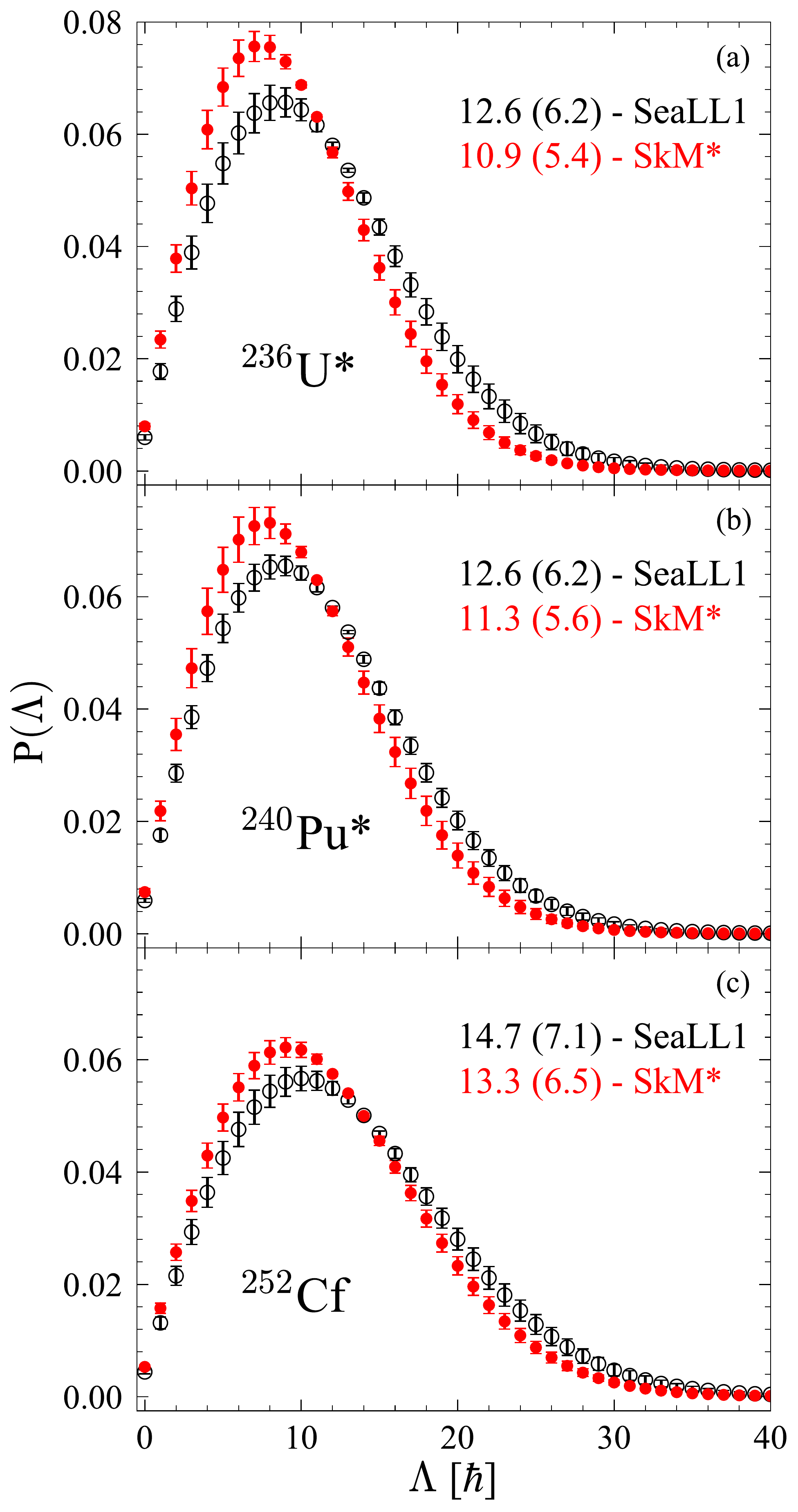}
\caption{ \label{fig:orbital} The orbital angular momentum distribution for three actinides.
For each nucleus 
the average and the corresponding standard deviations are shown in the legend.    The ``uncertainties'' are 
the standard deviations characterize the range of the variations due to 
the spread of the initial values of the multipole moments $Q_{20}$ and $Q_{30}$ and the energies of 
the fissioning nucleus~\cite{Bulgac:2016,Bulgac:2019b,Bulgac:2020,Bulgac:2021} and 
thus these distributions are characteristics for average FF splittings. }
\end{figure}

We performed TDDFT fission calculations of $^{236}$U, $^{240}$Pu, and $^{252}$Cf using two different nuclear energy density functionals  (NEDFs), 
SkM$^*$~\cite{Bartel:1982} and SeaLL1~\cite{Shi:2018}, in simulation boxes $30^2\times60$  with a lattice constant and
$l=1$ fm and a corresponding momentum cutoff $p_\text{cut}=\pi\hbar/l\approx 600$ MeV/c, and
using the LISE package as described in Refs.~\cite{Bulgac:2016,Shi:2020,Bulgac:2020}. 
The excitation energies for $^{236}$U and $^{240}$Pu were chosen close to the neutron threshold, 
thus emulating the reactions $^{235}$U(n$_\text{th}$,f) and $^{239}$Pu(n$_\text{th}$,f).
The  initial nuclear wave function $|\Phi\rangle$ was evolved in time from various initial deformations $Q_{20}$ and $Q_{30}$
of the mother nucleus near the outer saddle until the FFs 
were separated by more than 30 fm as in Refs~\cite{Bulgac:2019b,Bulgac:2020,Bulgac:2021} and their shapes relaxed. In the case of $^{252}$Cf(sf) 
we started the simulation outside the barrier for energies close to the ground state energy.
The current implementation of the TDDFT framework~\cite{Bulgac:2013a,Bulgac:2019}
has proven capable of providing answers to a wide number of problems in cold atom physics, quantum turbulence in 
fermionic superfluids, vortex dynamics in neutron star crust, nuclear fission, and reactions. 
Density Functional Theory and Schr{\" o}dinger descriptions are mathematically identical quantum many-body frameworks for one-body 
densities~\cite{Dreizler:1990lr,Gross:2006,Gross:2012}, with the proviso that in nuclear physics neither 
NEDF nor the internucleon forces are known with sufficient accuracy~\cite{Salvioni:2020}.

\begin{figure}
\includegraphics[width=1\columnwidth]{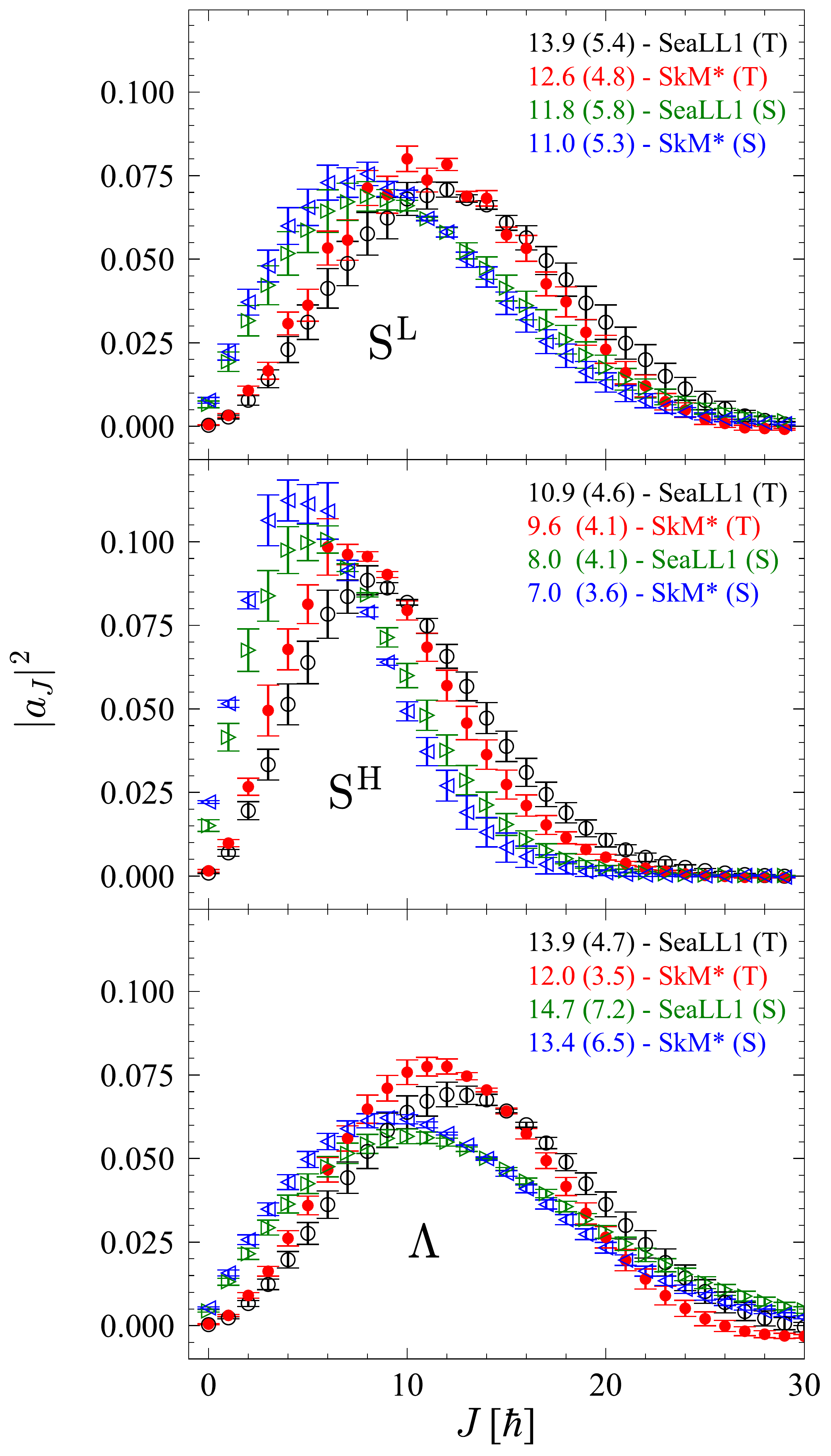}
\caption{ \label{fig:Uu} The light and  heavy  FF intrinsic spins and the orbital angular momentum 
distributions in case of spontaneous fission of $^{252}$Cf using the triple projection distributions from  Eqs.~(\ref{eq:L}-\ref{eq:Lam})~\cite{Bulgac:2021c} and from the single projection of 
the FF intrinsic spins~\cite{Bulgac:2021} and of the orbital angular momentum $\Lambda$ and  the corresponding   
average values for the intrinsic spin or the orbital angular momentum (standard deviation). 
(T) and (S) stand for the triple and single projections of the angular momenta. }
\end{figure}

\begin{figure}
\includegraphics[width=1\columnwidth]{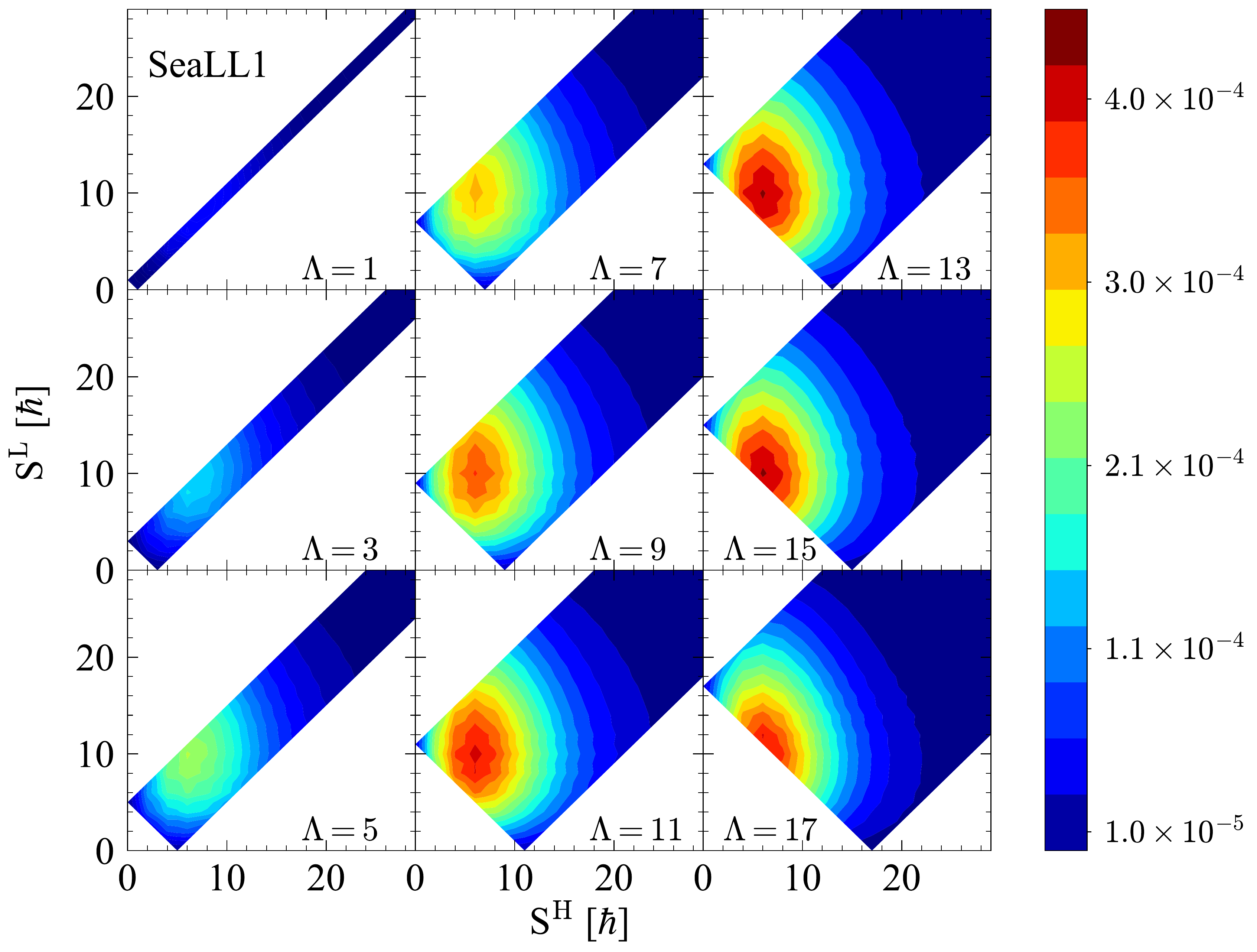}
\includegraphics[width=1\columnwidth]{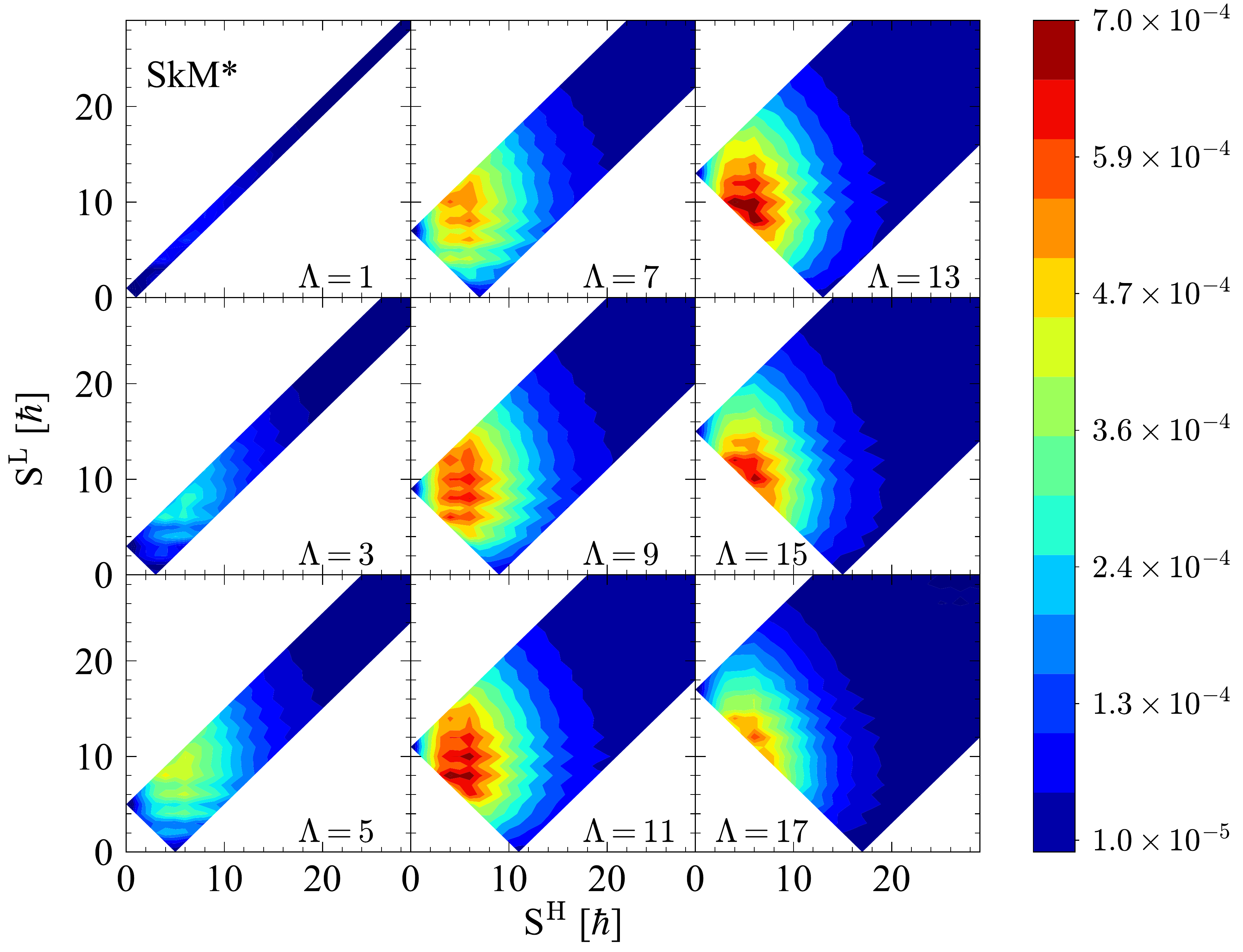}
\caption{ \label{fig:Le}The $^{252}$Cf triple probability distribution $P(\Lambda,S^\text{L},S^\text{H})$ for  SeaLL1 (upper panel) 
and SkM$^*$ (lower panel) NEDFs for odd values of $\Lambda$. 
The FF parities are correlated with the orbital angular momentum $\pi^\text{L}\pi^\text{H}=(-1)^\Lambda$. 
This triple distribution vanishes outside the region $|S^\text{L}-S^\text{H}| \le \Lambda \le S^\text{L}+S^\text{H}$, 
shown with white in these plots. The distributions for $^{236}$U$^*$ and $^{240}$Pu$^*$ are very similar.}
\end{figure}

\begin{figure}
\includegraphics[width=1\columnwidth]{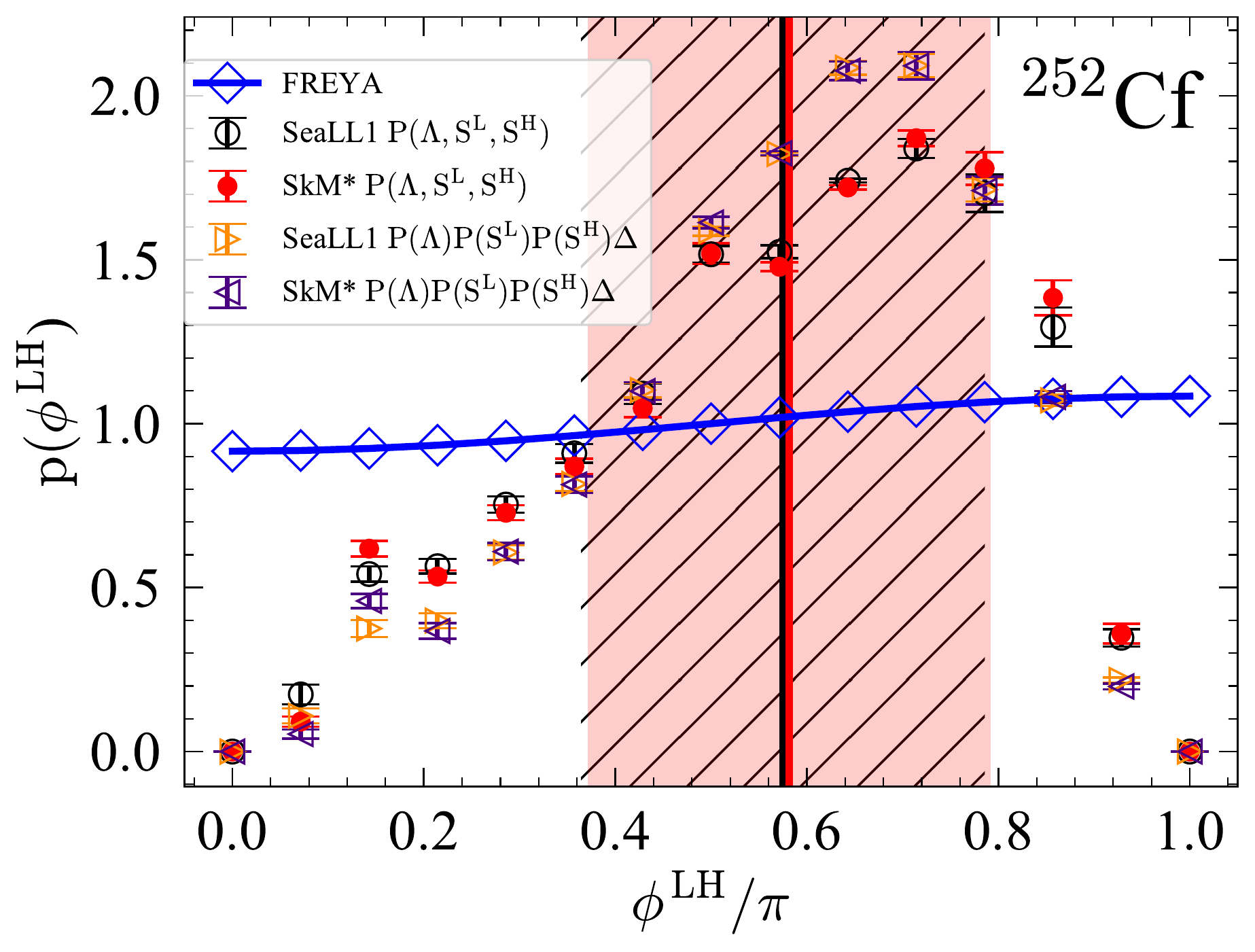}
\caption{ \label{fig:Cos} The  circles and bullets
represent the histogram (bin size = 0.22 radian) of the angle between the FF intrinsic spins ${\bm S}^\text{L}$ and ${\bm S}^\text{H}$, 
extracted using the triple distribution $P(\Lambda,S^\text{L},S^\text{H})$ and  Eq.~\eqref{eq:cos} to evaluate $p(\phi^\text{LH})$, 
$\int_0^\pi d\phi^\text{LH}p(\phi^\text{LH})=1$. 
The triangles represent the histogram obtained with $P(\Lambda)P(S^\text{L})P(S^\text{H})\Delta$, see text and Eqs.~(\ref{eq:L}-\ref{eq:Delta}). 
 The blue line and diamonds are the prediction of the FREYA model~\cite{Randrup:2021}. 
The distributions   $p(\phi^\text{LH})$ for $^{236}$U$^*$ and $^{240}$Pu$^*$ are very similar.}
\end{figure}

The distributions of the FFs' orbital angular momenta, see Fig.~\ref{fig:orbital}, 
are the first unrestricted microscopic extractions of these quantities.
As the masses of $^{236}$U,  $^{240}$Pu, and $^{252}$Cf are close to one another, the $\Lambda$ 
distributions  obtained by performing 
a single angular projection of the overlap $\langle \Phi | \Phi(\beta_0) \rangle  = \langle \Phi | e^{i\beta_0(J_x^\text{L}+J_x^\text{H})} |\Phi\rangle$, 
as in Ref.~\cite{Bulgac:2021}, are very similar.
Such individual intrinsic spin distributions can be recovered independently 
from our triple projection results from $P(\Lambda,S^\text{L},S^\text{H})$ as follows
\begin{align}
\!\!\!P(S^\text{L,H}) &= \sum_{ S^\text{H}\,\text{or}\, S^\text{L},\Lambda } \!\!\!P(\Lambda,S^\text{L},S^\text{H}), \, \sum_{S^{L,H}} P(S^\text{L,H}) = 1,\label{eq:L} \\
P(\Lambda) &= \sum_{ S^\text{L,H}}, \, \sum_\Lambda P(\Lambda) = 1, \label{eq:Lam}
\end{align}
and a comparison between results using the single and the triple projections 
in case of induced fission of $^{252}$Cf are shown in Fig.~\ref{fig:Uu}. The more precise triple projection leads to larger 
FF intrinsic spins by about $2\ldots 3\, \hbar$, while the average orbital 
angular momentum $\Lambda$ decreases by about 1 $\hbar.$ (Similar corrections 
to the FF intrinsic spins would be required for the estimates presented in Ref.~\cite{Marevic:2021}.)
As demonstrated in Ref.~\cite{Stetcu:2021}, 
the emission of neutrons and statistical gammas reduces the FF spins by $\approx 3.5-5\,\hbar$ by the time the FF decay 
reaches the yrast bandhead, corresponding to the FF spin values measured by \textcite{Wilson:2021}. 
The sum of the  yrast bandhead spins~\cite{Wilson:2021} for $^{252}$Cf, 
$6.85\,\hbar$ for the heavy FF and 6.44 $\hbar$ for the light FF respectively (averaged over all measured FFs)
with the angular momentum loss to decay $\approx 3.5-5\, \hbar$
estimated in Ref.~\cite{Stetcu:2021}, using standard phenomenological inputs however,  
agree reasonably well with our estimates of the average intrinsic FF spins in Fig.~\ref{fig:Uu}. 

In Fig.~\ref{fig:Le} we show the triple distribution $P(\Lambda,S^\text{L},S^\text{H})$ for
odd values of $\Lambda$. 
The even values of $\Lambda$ fixes both FF parities to be identical, $\pi^\text{L}=\pi^\text{H}$, 
while in case of odd $\Lambda$ these parities are opposite, $\pi^\text{L}=-\pi^\text{H}$, since for  $^{252}$Cf 
$S_0^\pi = 0^+$.  The distribution $P(\Lambda,S^\text{L},S^\text{H})$ is nonvanishing only in the region defined by Eq.~\eqref{eq:3Js}. 

The distribution of the angles between the intrinsic spins ${\bm S}^\text{L}$ and ${\bm S}^\text{H}$ is particularly instructive 
and qualitatively different from previous predictions. 
It was assumed a number of times in the literature, see Refs.~\cite{Randrup:2021,Vogt:2020} and references therein, 
that the two intrinsic spins are very weakly correlated at most. In particular, this was one of the main interpretations of the experimental results 
recently published by \textcite{Wilson:2021}. If that were case, the distribution $p(\phi^\text{LH})$ would basically 
be flat, similar to the predictions in Refs.~\cite{Vogt:2020,Randrup:2021}, with those results reproduced in this figure.
In Fig.~\ref{fig:Cos}, the distribution $p(\phi^\text{LH})$ 
evaluated by us is clearly not a uniform distribution, with a prominent maximum at an angle $\phi^\text{LH}\approx 2\pi/3$~\cite{Bulgac:2021d}. 
The probability of having angles $\phi^\text{LH}\ge \pi/2$ is $\approx 0.71/0.72$ (SeaLL1/SkM$^*$), which means that the two FF intrinsic 
spins are predominantly pointing in opposite directions and that the
the bending modes are predominantly favored over the wriggling modes. 
In Fig.~\ref{fig:Cos}, we used instead of 
the correlated evaluated distribution $P(\Lambda,S^\text{L},S^\text{H})$ the uncorrelated distribution $P(\Lambda)P(S^\text{L})P(S^\text{H})\triangle $ 
obtained using Eqs.~(\ref{eq:L}-\ref{eq:Lam}), shown with triangles. The results appear very similar, even though  $P(\Lambda,S^\text{L},S^\text{H})$  
is drastically different from $P(\Lambda)P(S^\text{L})P(S^\text{H})\triangle $, where 
$P(\Lambda)P(S^\text{L})P(S^\text{H})$ 
is non-vanishing outside the region~\eqref{eq:3Js}, the white regions in Fig.~\ref{fig:Le} and in evaluating the 
distribution $p(\theta_\text{LH})$  we have imposed the~\eqref{eq:3Js} restriction and renormalized the distribution
 $P(\Lambda)P(S^\text{L})P(S^\text{H})\triangle$ by a (not shown) factor $\approx 0.74$. 
 Fig.~\ref{fig:Cos} unfortunately does not reveal the large amount of FF intrinsic spin correlations, which are not merely geometrical in nature, since
 \begin{align}
\!\!\!\!\!\!\!\sum_{ S^\text{L,H},\Lambda } |  P(\Lambda)P(S^\text{L}) P(S^\text{H})\triangle  - P(\Lambda,S^\text{L},S^\text{H})| =0.35,
 \end{align}
 when the geometrical constraint Eq.~\eqref{eq:Delta} is taken into account.
  
In Fig.~\ref{fig:Cos} we plot the recent published results obtained with the phenomenological 
model FREYA, where \textcite{Randrup:2021} discussed the generation of the fragment angular momentum in fission. 
In Ref.~\cite{Randrup:2021} the claim is made that, unlike the conclusion reached by \cite{Wilson:2021}, i.e.,
that the FF intrinsic spins were formed after scission and are uncorrelated, the primordial intrinsic spins emerge uncorrelated before scission. 
This argument is based on the assumptions that the FF spins dynamics is governed by the rotational energy 
\begin{align}
E_\text{rot}=  
                     \frac{ {\bm S}^\text{L}\cdot {\bm S}^\text{L}} {  2I^\text{L} } + 
                     \frac{ {\bm S}^\text{H}\cdot {\bm S}^\text{H} }{ 2I^\text{H} }+\frac{{\bm \Lambda}\cdot {\bm \Lambda} }{2I^\text{R}},
                     \label{eq:rot}
\end{align}
where $I^\text{L,H,R}$ are the FFs and orbital moments of inertia, satisfying the relation $I^\text{R}\approx 10\, I^\text{L,H}$. 
The only correlation between $S^\text{L,H}$ is due to the third term, which is quantitatively small and  
which one can hardly quantify as highly correlated, is in stark contradistinction 
with our microscopic results in the same figure. While at first glance this assumption appears valid, see also Refs.~\cite{Moretto:1980,Vogt:2020}, 
upon closer analysis it becomes clear that the most general form allowed by symmetry is
\begin{align} 
E_\text{rot} = ({\bm S}^\text{L},{\bm S}^\text{H},{\bm \Lambda})^\text{T}  \otimes \tensor {\bm I}   
\otimes ({\bm S}^\text{L},{\bm S}^\text{H},{\bm \Lambda}),
\end{align} 
with a nondiagonal $3\times 3$  effective inertia tensor $\tensor{\bm I}$ in general.

The impact of the emission of neutrons and $\gamma$ rays on the spin of the FFs was discussed
in Ref.~\cite{Stetcu:2021} within the Houser-Feshbach framework~\cite{Hauser:1952}, where it was demonstrated that 
the intrinsic FF spins can be changed on average by 3.5 - 5 $\hbar$, a 
process that leads to a strong decorrelation of the observed FF spins, 
a process strongly underestimated by the model of Ref.~\cite{Wilson:2021}. The experimental data~\cite{Wilson:2021} characterizes 
only the yrast bandhead FF spins  after a large amount of the internal FF excitation energy, $\approx $ 20 MeV per 
FF~\cite{Bulgac:2016,Bulgac:2019b,Bulgac:2020,Schmidt:2018,Talou:2021,Randrup:2009}, 
was carried away by emitted particles.
The work presented here can better guide phenomenological models~\cite{Vogt:2020,Randrup:2021,Becker:2013,Talou:2021} 
and further extend the analysis in Ref.~\cite{Stetcu:2021}, which all rely on a quite large number of fitting parameters.

\begin{acknowledgements}

We want to thank G. Scamps for numerous comments. AB wants also to thank L. Sobotka for quite a number 
of discussions related to the role of FF intrinsic spin dynamics and older results in literature and reading an 
earlier version of the  manuscript.
AB was supported by U.S. Department of Energy,
Office of Science, Grant No. DE-FG02-97ER41014.   
The work of AB (partially)  and IA was  supported by the Department of 
Energy, National Nuclear Security Administration, under Award Number DE-NA0003841. 
KG was supported by NNSA Cooperative Agreement DE-NA0003885.
The work of IS was supported by the US Department of Energy through the 
Los Alamos National Laboratory. Los Alamos National Laboratory is operated 
by Triad National Security, LLC, for the National Nuclear Security Administration 
of U.S. Department of Energy Contract No. 89233218CNA000001. 
IS gratefully acknowledges partial support by the Laboratory Directed Research 
and Development program of Los Alamos National Laboratory under project 
No. 20200384ER and partial support and computational resources 
provided by the Advanced Simulation and Computing (ASC) Program.
This research used resources of the Oak Ridge
Leadership Computing Facility, which is a U.S. DOE Office of Science
User Facility supported under Contract No. DE-AC05-00OR22725 and of
the National Energy Research Scientific Computing Center, which is
supported by the Office of Science of the U.S. Department of Energy
under Contract No. DE-AC02-05CH11231.
This research also used resources provided by the Los Alamos National Laboratory 
Institutional Computing Program.
AB devised the theoretical framework. IA, KG, and IS performed TDDFT calculations, 
implemented, and performed the extraction of the spin distributions. 
All authors participated in the discussion of the results and the writing of the manuscript.  

\end{acknowledgements}


\providecommand{\selectlanguage}[1]{}
\renewcommand{\selectlanguage}[1]{}

\bibliography{latest_fission}

\end{document}